\documentclass[english,showpacs,floatfix,amsmath,amssymb,12pt]{revtex4}
\usepackage[dvips]{graphicx}
\usepackage{longtable}
\usepackage{amsmath,amssymb}
\usepackage{dcolumn}
\usepackage{latexsym}
\usepackage{babel}
\usepackage[latin1]{inputenc}
\usepackage{color}
\usepackage[colorlinks]{hyperref}
\newcommand\tabcaption{\def\@captype{table}\caption}
\def\al{\alpha}

\def\kp{\kappa}

\def\pa{\partial}
\def\vf{\varphi}
\def\vep{\varepsilon}

\def\om{\omega}
\def\Om{\Omega}

\def\be{\beta}
\def\dl{\delta}
\def\Dl{\Delta}

\def\La{\Lambda}
\def\th{\theta}

\def\diag{\mbox {diag}}

\def\ln{\mbox {ln}}
\def\wt{\widetilde}
\def\l{\left}
\def\r{\right}

\begin{document}
\title{\Large\bf Some Subtle Concepts in Fundamental Physics}
\author{Ying-Qiu Gu}
\email{yqgu@fudan.edu.cn} \affiliation{School of Mathematical
Science, Fudan University, Shanghai 200433, China} 
\date{20th July 2017}

\begin{abstract}
In this paper, we discuss some subtle concepts, such as coordinate,
measurement, simultaneity, Lorentz-FitzGerald contraction,
singularity in fundamental physics. The explanations of these
concepts in textbooks are usually incomplete and lead to puzzles.
Some long-standing paradoxes such as the Ehrenfest one are caused by
misinterpretation of these concepts. The analysis shows these
concepts all have simple and naive meanings, and can be well
understood using suitable logical procedure. The discussion may shed
light on some famous paradoxes, and provide some new insights into
the structure and features of a promising unified field theory.

\end{abstract}
\maketitle

\section{Coordinate vs. measurement}
\setcounter{equation}{0}

The effects of the special relativity such as length contraction and
time dilation of a moving object are still puzzling and
controversial problems, which result in some famous paradoxes like
the Ehrenfest one. To clarify the concepts of measurement and
observation, there were once some careful discussions, e.g. by
Terrell\cite{contr1}, Penrose\cite{contr2}, Weisskopf\cite{contr3}.
Under some approximation, these researches reached the result that
the photograph of a moving object will show it turns a little angle,
which is call the Penrose-Terrell rotation, but the length
contraction can not be recorded by the film\cite{contr7}. The
picture of a moving ball is still a ball. However, this problem is
not finally resolved, and some discussions still carry
on\cite{contr4,contr5,contr6}. As pointed out by M. Pardy in
\cite{contr6}, the investigation of relativity axioms as the
noncontradictory system was to our knowledge not published, although
we believe that the relativistic system of axioms is not
contradictory.

There maybe some basic concepts are confused and misused. In what
follows, we try to clarify this problem through detailed
calculation. To discuss the effects of relativity, we usually refer
to `measurement' and `observation', where the measurement actually
corresponds to the `coordinates', and `observation' to the process
of taking a snapshot of the moving object. In what follows we take a
ball, a screen and a curve moving with respect to a camera to show
their coordinates and the photographs.

Assuming $S(T, X ,Y, Z)$ is a stationary reference frame, and
$O(t,x,y,z)$ is a reference frame moving along $X $ at speed $V$
with respect to $S$, then we have the following Lorentz
transformation for coordinates
\begin{eqnarray}
T =t \cosh\xi +x\sinh\xi  ,\quad X = t\sinh\xi + x\cosh\xi ,\quad Y = y,\quad Z = z, \label{1.1}\\
t =  T \cosh\xi  - X\sinh\xi ,\quad x =  X\cosh\xi -
 T \sinh\xi,\quad y = Y,\quad z = Z, \label{1.2}
\end{eqnarray}
where $\tanh\xi= V$ stands for rapidity, and we set $c=1$ as the
unit of speed. (\ref{1.1}) and (\ref{1.2}) keep the 4-dimensional
length of a vector invariant
\begin{eqnarray}
T^2 -X^2- Y^2-Z^2 = t^2-x^2-y^2-z^2.\end{eqnarray} The coordinates
$(T, X ,Y, Z)$ and $(t,x,y,z)$ are mathematical maps which are
labeled to each event or 4-d point of space-time aforehand by
different researchers according to some rules.

 In newtonian space-time, we have a universal time $t$ and the
spatial reference frame has clear kinematic meaning. But in 4-d
Minkowski space-time, the time $t$ and spatial coordinates form a
quaternion entirety and can not be intuitively separated with each
other. So it is more natural to call the `reference frame' as
`coordinate system'. The 3-d kinematic explanation for a boosting
`reference frame' should be carefully treated, because the time $t$
is no longer universal now, which is closely connected with the
reference frame. The Lorentz transformation (\ref{1.1}) or
(\ref{1.2}) is more naturally regarded as a 1-1 correspondence
between two coordinate systems for each event. The event such as a
photon received by camera is an objective existence in the world,
but its coordinate, the photon is received at moment $T$ and place
$(X,Y,Z)$, is subjective label which is related to concrete
coordinate system preset by specific researcher.

At first, we consider a self-luminous sphere of radius $R$ located
at the origin of the static coordinate system $S$. Then an idealized
stereographic camera can record each point of the sphere. The
equation of the sphere is described by
\begin{eqnarray}
 X^2 + Y^2 +Z^2 =R^2, \label{1.3}
\end{eqnarray}
with any constant $T = T_0$. Substituting (\ref{1.1}) into
(\ref{1.3}) we get the equation of the sphere in the moving system
$O$ as
\begin{eqnarray}
\frac {(x+Vt)^2}{1-V^2}+y^2+z^2=R^2. \label{1.4}
\end{eqnarray}
So for any given moment $t$ in $O$, the sphere becomes an
ellipsoid due to the Lorentz-FitzGerald contraction. (\ref{1.4})
is only related to mathematical concept `coordinate', but
independent of physical process.

\begin{figure}
\centering
\includegraphics[width=17cm]{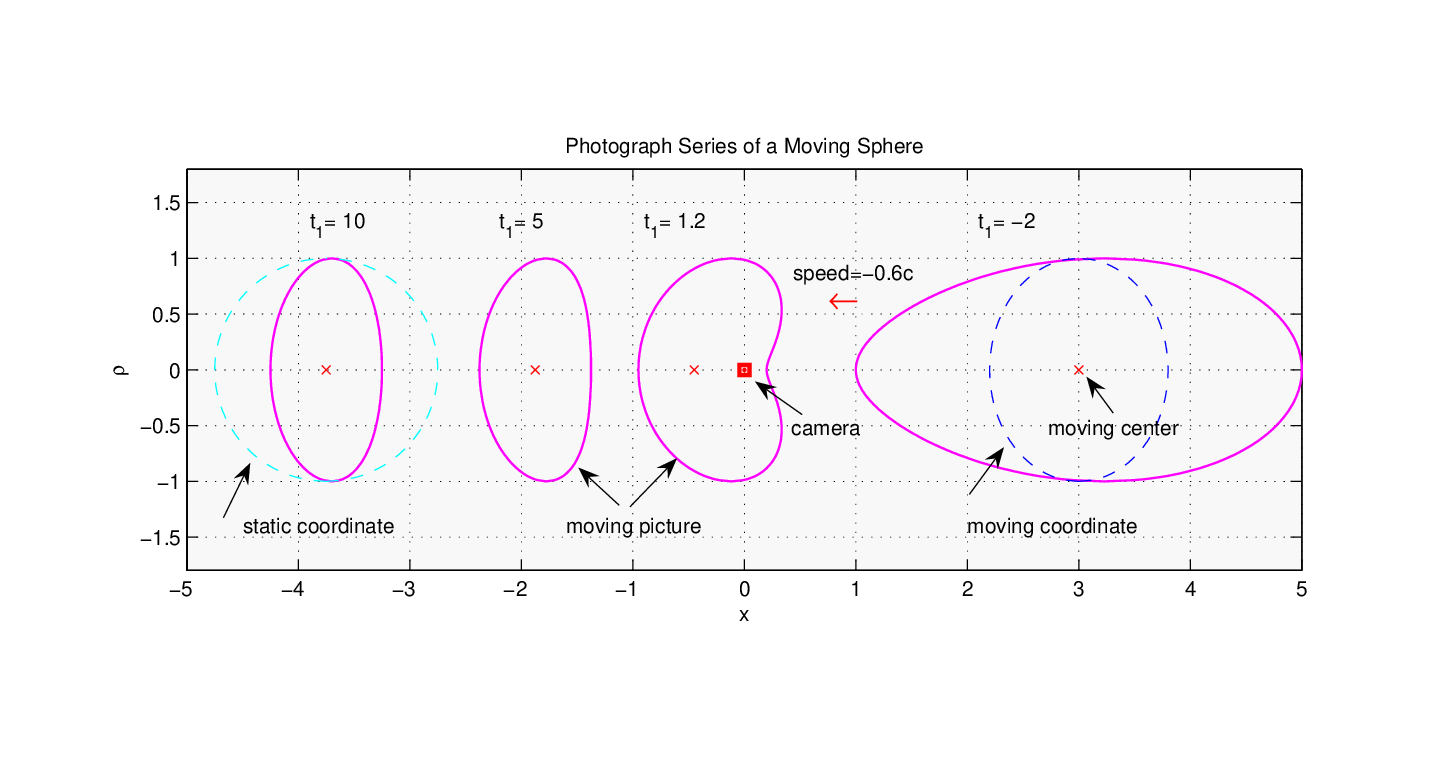}
\caption{The photograph series of a unit sphere recorded by the
camera} \label{fig1}
\end{figure}

Now we take a photograph of the sphere, which involves the physical
process of propagation of photons. Setting a camera at the origin of
the moving coordinate system $O$, it takes a snapshot of the sphere
at the moment $t_1$, then the photons received by the film are
emitted from $\vec r =(x,y,z)$ at time $t=t(\vec r)$, where $t(\vec
r)$ depends on the point $\vec r$. Note the velocity of the light is
constant $c=1$, so we have relation
\begin{eqnarray}
t=t_1-\frac r c=t_1-\sqrt{x^2+\rho^2},\qquad \rho\equiv
\sqrt{y^2+z^2}.
 \label{1.5}
\end{eqnarray}
In (\ref{1.5}), the coordinate of the sphere is constrained by
geometric equation (\ref{1.4}). Substituting  (\ref{1.5}) into
(\ref{1.4}) we get
\begin{eqnarray}
{\frac { [ x+V ( {t_1}-\sqrt {{x}^{2}+{\rho}^{2}}
 )  ] ^{2}}{1-{V}^{2}}}+{\rho}^{2}={R}^{2}
, \label{1.6}
\end{eqnarray}
where $(x,\rho)$ is the coordinate of the sphere record by the
film at moment $t_1$, and $t_1$ is a constant. At different moment
$t_1$, we get different picture of the sphere. The profiles
recorded by the picture are displayed in FIG.(\ref{fig1}), which
are quite different from the original sphere. The coming sphere
looks like an olive, but the departing sphere looks like a disk,
which contracts heavier than the Lorentz-FitzGerald contraction.

Setting a screen orthogonal to the $X$-axis, then the coordinates
of the screen is given by
\begin{eqnarray}
X \equiv X_0 = t\sinh\xi + x\cosh\xi . \label{1.7}
\end{eqnarray}
Substituting (\ref{1.5}) into (\ref{1.7}), we get the picture of
the screen, which becomes a standard hyperboloid,
\begin{eqnarray}
{\frac {(x+a)^{2}}{\sinh^2\xi}}-{ \rho}^{2}= ( X_0 -t_1 \sinh \xi)
^ {2},\qquad a = (X_0- t_1 \sinh\xi)\cosh\xi. \label{1.8}
\end{eqnarray}
Indeed, the picture of a piece screen with $\rho>0$ will rotate an
angle. However, the angle is not a constant, which depends on the
radial distance $\rho$, speed $V$ and $x_0$.

In the case of a screen orthogonal to the $z$-axis, namely,
$z=z_0$, the picture is still a plane. However, the curves in the
screen are distorted, so we consider an arbitrary spatial curve in
the static coordinate system $S$. The curve is described by
\begin{eqnarray}
X= X(\zeta),\quad Y= Y(\zeta),\quad Z= Z(\zeta),
\label{1.10}\end{eqnarray} where $\zeta$ is parameter of the
curve. Substituting (\ref{1.10}) into (\ref{1.2}), and then
inserting the results into (\ref{1.5}), we can solve
\begin{eqnarray}
T(\zeta)= t_1\cosh\xi
-\sqrt{(X-t_1\sinh\xi)^2+Y^2+Z^2},\label{1.11}\end{eqnarray} which
is the time in $S$ between the photons are emitted from point
$\zeta$ and received by the film. Again inserting (\ref{1.11}) into
(\ref{1.2}), we get the coordinates of the point $\zeta$ recorded by
the film
\begin{eqnarray}
x= X(\zeta)\cosh\xi-T(\zeta)\sinh\xi,\quad  y= Y(\zeta),\quad z=
Z(\zeta).\end{eqnarray} For a rigid bar lie at $Y=Y_0,Z=0$, we
have
\begin{eqnarray}
x=
(X-t_1\sinh\xi)\cosh\xi+\sqrt{(X-t_1\sinh\xi)^2+Y_0^2}\sinh\xi.\end{eqnarray}
It is still quite complex. We further assume $Y_0=0,X\in[-L,L]$
and $L>t_1\sinh\xi$, then we have
\begin{eqnarray}
x\in [ -(L+t_1\sinh\xi)\frac {1-V}{\sqrt{1-V^2}},
(L-t_1\sinh\xi)\frac {1+V}{\sqrt{1-V^2}}].\end{eqnarray} When
$t_1\to0$, the length of the bar `we see' becomes
$2L(1-V^2)^{-\frac 1 2}>2L$, which is not contracted but even
expanded. Similar to the moving sphere, the coming part of the bar
looks expansion, but the departing part looks contraction.

The above examples show how the concepts of special relativity work
consistently in logic. The point of view is simple and direct: The
coordinates $(T, X ,Y, Z)$ and $(t,x,y,z)$ are just labels of
mathematical map rather than realistic physical observation, but
they are language to express events and observation. The Lorentz
transformation (\ref{1.1}) or (\ref{1.2}) is nothing but a 1-1
mapping rule between two label systems for each event in Minkowski
space-time, which is purely geometric rather than kinetic and keeps
the metric $\eta_{\mu\nu}=\diag(1,-1,-1,-1)$ invariant. The special
relativity reveals the connection between time and space, but they
are independent in the Newtonian world.

The moment $t_1$ of taking snapshot is related to the physical
process of light transmission. A static camera and a moving one at
the same place at a moment to take snapshot will receive different
photons from the object, so the pictures look quite different if $V$
is large enough.

\section{Equations vs. Reality}
\setcounter{equation}{0}

All physical laws are represented by equations and relations, but a
realistic physical process is just one solution of these
equations\cite{Ellis3}. In some sense, our symbolic system is much
larger than the realistic world. Nevertheless, some related concepts
are usually confused. Now we examine the simultaneity in relativity.

\begin{figure}
\centering
\includegraphics[width=17cm]{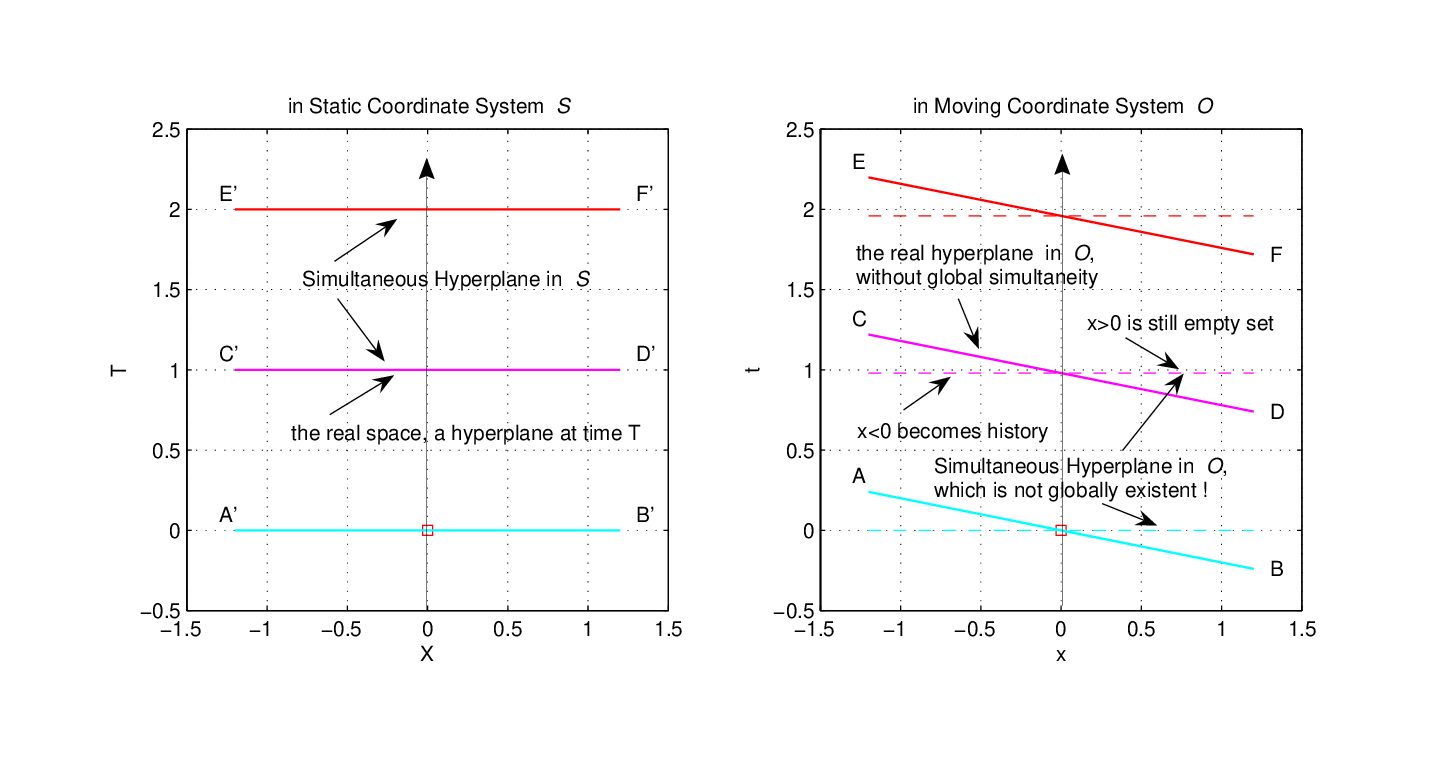}
\caption{The realistic space of the world is an evolving hyperplane.
Only in one special coordinate system we have global simultaneity.}
\label{fig2}
\end{figure}

Again we consider two coordinate systems of Minkowski space-time
with relative speed $V>0$, that is, $S$ and $O$ with the coordinate
transformation (\ref{1.1}) and (\ref{1.2}).  We examine the whole
world in coordinate system $S$. The world and the space-time are
evolving, and at time $T=0$, we have a map of the space
simultaneously, which is a hyperplane corresponding to the line
${A'B'}$ in FIG.(\ref{fig2}). The hyperplane $A'B'$ evolves into
$C'D'$ at time $T=T_0$ and into $E'F'$ at time $T=T_1$.  The global
simultaneity in $S$ means events at different places can
realistically occur at the same moment $T\equiv T_k$, rather than we
can take coordinate $T=T_k$ everywhere in theoretical sense. The
corresponding hyperplanes in the moving coordinate system $O$ are
$AB, CD$ and $EF$ respectively. The evolution of the hyperplanes
forms an evolving universe\cite{Ellis1,Ellis2}.

From FIG.(\ref{fig2}) we find that, only in one special coordinate
system, the realistic world can define the concept of global
simultaneity, otherwise it will result in contradiction. We show
this by detailed discussion. Assume that the global simultaneity
exists in coordinate system $S$, that is the world evolves from
$T=0$ to $T=T_0$ as shown in FIG.(\ref{fig2}). In this case,
according to the Lorentz transformation (\ref{1.2}), the realistic
space evolves from the hyperplane $AB$ to $CD$ in $O$, which is a
tilting line without simultaneity. If we toughly define a
simultaneity $t\equiv t_0$ in system $O$ and believe it can be
realized, we find that we actually erase the history in the region
$x<0$, and fill up the future in the region $x>0$. Of course such
situation is absurd, because the evolution of the world is even not
uniquely determined, and we can neither exactly forecast the future
nor change the history.

We examine the contradiction more clearly by the following events.
The first event $E_1$ is a photon emitted at moment $T=0$ and place
$(X,Y,Z)=(-1,0,0)$ in $S$, and the second event $E_2$ is another
photon emitted at moment $T=0$ and place $(1,0,0)$, and the third
event $E_3$ is the two photons are received simultaneously by a
camera located at the origin $(0,0,0)$ at the moment $T=1$. Of
course the events $E_1$ and $E_2$ take place simultaneously in $S$.

In the moving coordinate system $O$, by the Lorentz transformation
(\ref{1.2}), we have the coordinates for the three events
respectively,
\begin{eqnarray}
t =  \sinh\xi>0 ,\quad (x,y,z) =  (-\cosh\xi,0,0),\quad {\rm
for}~~E_1
\label{e1}\\
t =  -\sinh\xi <0,\quad (x,y,z) =  (\cosh\xi,0,0),\quad {\rm
for}~~E_2
\label{e2}\\
t =  \cosh\xi >0,\quad (x,y,z) =  (-\sinh\xi,0,0).\quad {\rm
for}~~E_3 \label{e3}
\end{eqnarray}
{ At moment $t=0$ in $O$, by (\ref{e2}) we find $E_2$ has taken
place, but by (\ref{e1}) $E_1$ does not occur yet. If we accept the
realistic simultaneity also exists in coordinate system $O$, i.e.,
events can take place simultaneously in $O$, then we can take off
the light at moment $t=0$ to cancel the first event $E_1$, and then
the third event $E_3$ will also be changed. Of course $E_1$ can not
be canceled, because it has actually occurred in the realistic
world. So the realistic simultaneity in coordinate system $O$ is
impossible.}

{ By Lorentz transformation (\ref{1.2}), we can solve
\begin{eqnarray}
t = \frac{T- x\sinh\xi}{\cosh\xi}. \label{tT}
\end{eqnarray}
For any simultaneity $T\equiv T_k$, (\ref{tT}) shows $t$ is not an
independent variable, which is determined by place $(x,y,z)$. This
also disproves the global simultaneity $t\equiv t_0$ exist in
coordinate system $O$.}

{ In theoretical sense, the hyperplanes $T\equiv T_0$ and $t\equiv
t_0$ describe two different possible worlds.  They are all
reasonable and all coordinate systems are equivalent. But in the
realistic world, we can define the simultaneity only in one class of
reference frames, i.e., the one is relatively static to the Cosmic
Microwave Background Radiation(CBMR). That is to say, the
equivalence of coordinate systems is just related to physical laws,
rather to their solutions. }

{ The difficulty in comprehension of principle of relativity is
caused by the fact that a strict definition or a standard
explanation for the principle is absent. The meaning of the
principle of relativity in Minkowski space-time is a little
different from that of Galileo's original thought experiment.
According to Galileo, in a closed ship, we cannot detect the uniform
motion of the ship. This thought experiment has a precondition:
there exist a universal $t$ and simultaneity. However, this
precondition is invalid in Minkowski space-time as shown above. In
Minkowski space-time, an elementary particle can detect its speed
relative to CMBR by compression its proper time $\tau$ with that of
other moving particles, and the particles with the largest $\tau$
are static to CMBR. So in this sense, not all inertial frames are
equivalent.}

{ The core meaning of principle of relativity is the covariance of
physical laws, that is, any physical law for a specific physical
system derived by different researchers is equivalent. Via a 1-1
regular transformation of coordinate system and physical variable
set, the physical law and their solutions derived by different
researchers can be transformed.  Of course, covariance includes two
respects, that is, the general coordinate system transformation and
the tetrad transformation. The Galileo's ship and the Einstein's
lift are great thought experiments to direct us to establish the
principle of covariance. However, the principle of covariance is an
extrapolation of the two thought experiments, and it is more
profound and accurate than the two thought experiments. So in
theoretical analysis, it is better to directly start from principle
of covariance, and then the confusion in logic can be avoided. }

The key of the problem is that, the realistic world is just one
solution of the dynamical equations. The principle of relativity and
the covariance of physical equations just mean they are suitable for
all cases of the evolving world, and identical for all researchers.
Of course, this feature is also consistent with causality, because
any space-like hyperplane in one coordinate system is space-like in
other ones, and any suitable initial data given in one hyperplane
suitable for the dynamical equations in one coordinate system are
also suitable for the dynamical equations in other ones. This is an
astonishing but natural characteristic of truths. A fundamental
postulate of the unified field theory should have such
characteristic.

In curved space-time, the realistic world also has unique unified
cosmic time and a simultaneous hypersurface related to the time
coordinate. The global simultaneity is just valid in the coordinate
system relatively static to the Cosmic Microwave Background
Radiation, which forms the uniform cosmic time\cite{quat}. So the
covariance of the physical laws does not contradict the existence of
a special coordinate system for the realistic world.

The lack of universal simultaneity is related to the approximation
of the classical mechanics\cite{gu1}, because the classical energy
and momentum of a particle are defined as the spatial integrals of
some N\"other's charges, which require the hyperplane with
simultaneity. The electric charge is special, which is exactly
conserved in all coordinate system for each spinor due to gauge
invariance, but the momentum of the spinors can be exactly
calculated in all coordinate system only if the spinors take energy
eigenstate, because in this case the N\"other's charge becomes time
independent, otherwise the exchange of energy-momentum with
environment exists. The Lorentz transformation, the time dilation
and length contraction are just a local effects. The proper time of
a moving particle is $\tau=\int_0^t\sqrt{1-v^2(t)}dt$, even if the
motion is accelerating.

The misuse of simultaneity will lead to  paradoxes. Now we consider
the famous Ehrenfest paradox in rotational cylindrical coordinate
system\cite{Ehren}. The transformation between Born chart
$B(t,r,\phi,z)$ and the static Cartesian chart $S(T,X,Y,Z)$ is given
by
\begin{eqnarray}
T=t,\quad X=r\cos(\phi+\om t),\quad Y= r\sin(\phi+\om t),\quad
Z=z. \label{2.3}\end{eqnarray} The line element in $B(t,r,\phi,z)$
becomes
\begin{eqnarray}
ds^{2}=( 1-{r}^{2}{\om}^{2} )  dt^{2}-2{r}^{2}\om dt d\phi
-dr^{2}-{r}^{2} d\phi^{2}-dz^{2}. \label{2.4}\end{eqnarray} It
includes the `cross-terms' $dt d\phi$, so the time-like vector
$\pa_t$ is not orthogonal to the spatial one $\pa_\phi$. This is
the key of the paradox. In the opinion of the observer in static
coordinate system $S$, the simultaneity means $dT=0$, which leads
to $dt=0$, and then the spatial length element is given by
\begin{eqnarray}
dl=|ds|_{dT=0}=\sqrt{dr^2+r^2d\phi^2+dz^2},
\label{2.5}\end{eqnarray} which is identical to the length in the
static cylindrical coordinate system. However, the temporal length
is different. For a clock attached to the rotational coordinate
system, we have $dr=d\phi=dz=0$, and in this case we should have
$r|\om|<1$. Then the proper time element is given by
\begin{eqnarray}
d\tau = |ds|_{dl=0}=\sqrt{1-{r}^{2}{\om}^{2}}dt,
\label{2.6}\end{eqnarray} which shows the relativistic effect, the
moving clock slows down. In the region $r|\om|\ge 1$, the Born
coordinate is still a 1-1 smooth mapping, so it is a valid
coordinate transformation in mathematics, but we can not fix a clock
in the rotational system $B(t,r,\phi,z)$ due to $ds$ being
imaginary, which means $dt$ becomes space-like element. This is one
difference between mathematical coordinate and physical process.

In the opinion of an observer attached at $(r_0,\phi_0,z_0)$ in
the rotational coordinate system, the spatial vector bases
$(\pa_r,\pa_\phi,\pa_z)$ are also orthogonal to each other, and
then the line element between two local points should take the
form $\dl s^2=\dl t^2-\dl r^2- g_{\phi\phi}\dl\phi^2-\dl z^2$,
where $\dl t$ is his local time element. By the universal
expression of line element (\ref{2.4}), we have
\begin{eqnarray}
ds^{2}=\dl t^{2}-dr^{2}-\frac {r^2}{1-r^2\om^2} d\phi^{2}-dz^{2},
\label{2.7}\end{eqnarray} where $r=r_0$ for this specified
observer and $\dl t$ is given by
\begin{eqnarray}
\dl t = \sqrt{1-r^2\om^2}dt-\frac
{r^2\om}{\sqrt{1-r^2\om^2}}d\phi. \label{2.8}\end{eqnarray} The
simultaneity of this observer means $\dl t|_{r=r_0}=0$. For all
observer attached in the rotational coordinate system, the induced
Riemannian line element in the quotient spatial manifold $(r,\phi,
z)$ is generally given by
\begin{eqnarray}
dl^{2}=dr^{2}+\frac {r^2}{1-r^2\om^2} d\phi^{2}+dz^{2},\quad
(r|\om|<1), \label{2.9}\end{eqnarray} which corresponds to the so
called Langevin-Landau-Lifschitz metric. However, the explanations
of the metric in textbooks are usually quite complicated.

Since the 1-form (\ref{2.8}) is not integrable, so we can not
define a global time for all observers attached in the rotational
coordinate system. The physical reason is that the underlying
manifold of the Born chart is still the original Minkowski
space-time. However, if we toughly redefine a homogenous time $\wt
t$ orthogonal to the hyersurface (\ref{2.9}), then we get a new
curved space-time equipped with metric
\begin{eqnarray}
ds^{2}=d {\wt t}^{~2}-dr^{2}-\frac {r^2}{1-r^2\om^2}
d\phi^{2}-dz^{2},~~(r|\om|<1). \label{2.10}\end{eqnarray} Needless
to say, (\ref{2.10}) and (\ref{2.7}) describe different space-time
and are not globally equivalent to each other, but they are
equivalent locally. So generally speaking, the Lorentz
transformation is only a local manipulation in the tangent
space-time of a manifold, and the global Lorentz transformation
only holds between Cartesian charts in the Minkowski space-time.
The paradoxes usually arise from that we are unaccustomed to
calculating 4-dimensional geometry.

The Rindler coordinate system is another example. The
transformation between the Rindler chart $R(t,x,y,z)$ and the
static Cartesian chart $S(T,X,Y,Z)$ of Minkowski space-time is
given by\cite{Rindler}
\begin{eqnarray}
T=x\sinh(t),\quad X= x\cosh(t),\quad Y=y,\quad Z=z,
\label{2.1}\end{eqnarray} which is a 1-1 smooth map in the Rindler
wedge $\{X\in(0,\infty), |T|<X, (Y,Z)\in R^2\}$. In the Rindler
chart the line element of the space-time becomes
\begin{eqnarray}
ds^2= x^2 dt^2- dx^2-dy^2-dz^2,\quad x\in(0,\infty),~(t,y,z)\in R^3.
\label{2.2}\end{eqnarray} The above transformation is well defined
in mathematics. The problem comes when making simultaneous kinematic
calculation. If we acquiesce the hyperplane $T=T_0$ in Cartesian
chart describe the realistic world, most part of the hyperplane
$t=t_0$ is still an empty set in physics as shown in
FIG(\ref{fig2}). The realistic space only corresponds to the
hypersurface $x\sinh(t)\equiv T_0$(constant), but the hyperplane
$t=t_0$ is another space-time. So the spatial length element
$dL=|ds|_{dT=0}$ can not be confused with $dl=|ds|_{dt=0}$.
Accordingly, the Rindler horizon is a meaningless concept in
physics. Notice that the Raychaudhuri equation is based on the
simultaneity of time-like congruence of the world lines, so it only
holds locally and is not suitable for discussion of the large scale
structure of the space-time.

The distinction between symmetry of dynamical equations and
solutions is also puzzling in field theories. The parity symmetry of
the Schr\"odinger equation or Dirac equation does not means their
solutions are certainly even or odd functions of the coordinates,
and the rotation symmetry of the equations also does not mean all
solutions are spherically symmetric. Even the eigen solution for a
single spinor always has a polar axis. We can never establish a
coordinate system  such that the solution of a molecule ${\rm NH_3}$
is of parity symmetry or rotation symmetry, but all such symmetries
certainly hold for its total Lagrangian. The permutational symmetry
of the many-body Schr\"odinger equation together with the Pauli
exclusion principle also does not mean the solutions should take the
form of Slater determinant. The symmetries of equations and
solutions are different, although there are some relationship
between them via Lie algebra\cite{Nobel}.

\section{Dirac delta vs. Singularity}
\setcounter{equation}{0}

In classical mechanics and electromagnetism we have an idealized
model, the Dirac-$\dl$, to describe a point mass or point charge,
which is just a mathematical abstraction of  relatively concentrated
distribution. Some corresponding singularities, such as infinity of
mass density or infinite field intension, can be easily understood
as a result caused by the simplification of the models. Even so,
some mathematical calculation is still not evident. For example,
\begin{eqnarray}
\Dl \frac x {r^3} ={4\pi}\frac \pa {\pa x} \dl(\vec r) \label{3.1}
\end{eqnarray}
is easily confused with $\Dl \frac x {r^3} =0,(r>0)$. Only by
using the test function $f(\vec r)\in C_c^\infty$ and calculating
the integral $\int _{R^3} f(\vec r) \Dl\frac x {r^3} dV$, we will
find out a source at the center and get the right relation
(\ref{3.1}).

In the nonlinear theories such as general relativity, the issue
becomes even ambiguous, because we have not a universal definition
of generalized function for nonlinear partial differential
equations. We take Schwarzschild space-time
\begin{eqnarray}
ds^2=b(r) dt^2- a(r) dr^2 -r^2( d\th^2 + \sin^2\th d\vf^2),
\label{1.12}
\end{eqnarray}
as an example to show the problem. In the case of standard
solution $b=a^{-1}=1-\frac {R_s} r$, the usual opinion believes
that the singularity on the horizon $r=R_s\equiv 2Gm$ is just a
coordinate singularity, and it can be removed by suitable
coordinate transformation. The {Kruskal coordinate is the most
popular one},
\begin{eqnarray}
  \tanh \frac t 2 &=& \left \{ \begin{array}{l}
    \frac \tau \rho, ~~r>R_s, \\
    \frac \rho \tau , ~~r \le R_s,
  \end{array} \right.\quad (r-R_s)\exp\l(\frac r{R_s}\r) =R_s(\rho^2-\tau^2).\label{1.13}\\
  ds^2&=& \frac {4R_s^3} r \exp\l(\frac r{R_s}\r) ( d \tau^2- d \rho ^2) -
  r^2 (d \theta ^2+ \sin^2 \theta d\varphi^2).
\label{(1.14)}\end{eqnarray} However, the transformation
(\ref{1.13}) may be invalid globally, because it even has not the
first order derivatives at $r=R_s$. In the viewpoint of differential
geometry, if $r=R_s$ has not singularity, the space-time in the
neighborhood is smooth manifold, and then the consistent local
coordinate transformation should be $C^\infty$. In the viewpoint of
partial differential equation, the Einstein field equation includes
second order derivatives, so a valid coordinate transformation
should at least has bounded second order derivatives. In the
viewpoint of logic, the exterior Schwarzschild solution has nothing
to do with the interior one, except that they are expressed by the
same Latin characters.

As a matter of fact, $r=R_s$ may be not only a singular surface,
but also a surface with concentrated mass-energy distribution. We
make some analysis. Consider the metric generated by a spherical
mass shell with thickness $\vep$ and outside radius $R$. The
equation governing $a$ is given by
\begin{eqnarray}
\frac d {dr} H(r)= 8\pi G \rho r^2, \qquad R_s = \int_{R-\vep}^R
8\pi G \rho r^2 dr, \label{1.15}
\end{eqnarray}
in which we define $H=r(1-a^{-1})$ in order to make a linear
equation, $\rho$ is the gravitating mass-energy density. Let
$\vep\to 0$, we get
\begin{eqnarray}
\frac d {dr} H(r)= R_s\dl(r-R), \label{1.16}
\end{eqnarray}
and the solution is given by
\begin{eqnarray}
a &=& \left \{ \begin{array}{ll}
1, &~~r<R, \\
\l(1-\frac {R_s} r\r)^{-1}, &~~r \ge R\ge R_s.
\end{array} \right.
\label{(1.17)}\end{eqnarray} The interior space-time is
Minkowskian, and the exterior space-time is Schwarzschild one.
Just like the case of (\ref{3.1}), the concentrated source is
sometimes not evident in the equations, especially for the
nonlinear differential equations.

Of course, (\ref{1.16}) is an extremely simplified model. A
realistic model should include the dynamical behavior of the
source. The calculation in \cite{gu2,gu3} shows the center of a
star is not a balance point of the particles, so the singular
surface $r=R_s\ne0$ may just reflect the intrinsic harmonicity of
the Einstein's field equation.

How to define the singularity in the space-time is a difficult
problem. The usual definitions are `incomplete geodesic' or infinite
scalar of curvatures such as $R\to \infty$, $R_{\mu\nu}R^{\mu\nu}\to
\infty$, $R_{\mu\nu\al\be}R^{\mu\nu\al\be}\to \infty$,
etc\cite{Hawking,Wald}. As a physical description, the incomplete
geodesic is rather difficult for manipulation. Besides, for an
evolving space-time, it is not a good choice to use the future
property of the manifold, because  we can not strictly forecast the
future as analyzed above. The singular curvatures also have a
disadvantage, that is, it can be easily overlooked or confused with
coordinate singularity like the case of Schwarzschild metric. The
most convenient definition maybe directly use the physical
singularity $\rho \to \infty$. Although it is equivalent to $R\to
\infty$ in mathematics, they are different in physics. This is
because, on the one hand, $\rho\to \infty$ has manifest physical
meaning and which can not be overlooked, but $R\to \infty$ is easily
overlooked similarly to $\Dl \frac 1 r =0,(r>0)$. On the other hand,
$\rho\to \infty$ occurs before the metric itself becomes
singular\cite{gu2,gu3}, so the space-time can be still treated as a
`place'\cite{Wald} when the mass-energy density approaches to
infinity.

\section{Energy vs. Entropy}
\setcounter{equation}{0}

In physics we have not only primary concepts, such as space-time
and field, but also have some profound secondary concepts helping
us to understand how the Nature works. The `Energy-momentum' and
the `Entropy' are the most important secondary ones. However, the
situations for the two concepts are quite different. The
energy-momentum of any system can be clearly defined from the
dynamical equations according to the N\"other's theorem. Its
general validity and corresponding conservation law are rooted in
the underlying principle of covariance. Even in the case that the
detailed dynamical equations of a complex system such as a
thermodynamic system is unknown, we can also well understand and
measure the total energy of the system and the exchange with the
environment.

The entropy is also used as a general concept for all system.
Indeed, for some idealized statistical models, we have clear
definition of entropy\cite{greiner}.  In the case of a reversible
thermodynamic process, we have a state function entropy calculated
by $dS= \dl Q/T$, where $\dl Q$ is the heat increment received by
the system, and we have the energy conservation law or Gibbs-Duhem
relation
\begin{eqnarray}
d E= \dl Q + \dl W= T dS- P dV+\sum_k \mu_k d N_k. \label{1.9}
\end{eqnarray}
But the definition only holds for idealized reversible process. In
the case of equilibrium idea gas with $N$ particles and volume
$V$, we can derive the elegant Sackur-Tetrode equation
\begin{eqnarray}
S= Nk\l[\frac 5 2 +\ln\l\{\frac V{Nh^3}\l(\frac{4\pi
mE}{3N}\r)^{\frac 3 2}\r\}\r]. \label{1.10*}
\end{eqnarray}
But it is also only valid for equilibrium state with some
assumptions. Furthermore, we still have more general Boltzmann
relation $S= k\ln\Om$, where $\Om$ is the number of microstate.

We can even prove all such definitions of entropy are consistent,
but unfortunately, there is no method to generalize these
statistical definitions to describe a system of internal structure
and organization. Different from the energy-momentum, the entropy
have not an underlying dynamical principle, although we usually
take the second thermodynamic law as a generally valid principle.
We find it is actually difficult to understand and to measure
entropy for any complex system. The order or disorder of a complex
system, the physical essence of entropy, is also an ambiguous
statement. Besides the statistical system, in the Nature there are
a lot of systems of internal structure, interaction and
organization. For example, we put some eggs in a black box and
then set the box in a heat bath with suitable temperature. After
some time, we will get alive chickens. If the temperature is
higher, we will get cooked eggs. How to measure and compare the
entropy of the two processes?  There are also the contrary
processes. Press some vapor into the black box and set it in a
cold environment, then we will get beautiful hexagonal snow, whose
molecules are automatically organized. So the concept entropy for
a complex system is quite ambiguous, and even can not be
understood in principle.  The misuse of such concept is usually to
make puzzles rather than to clarify any physical rule.

We usually use the second thermodynamic law to explain the arrow
of time. In this case it is somewhat similar to putting the cart
before the horse as pointed out by G. Ellis\cite{Ellis2}. The one
direction of time is a basic fact and characteristic of the
evolving world, so what we should do is to explain why our
linearized and simplified time-reversible theories are so
successful, and to find out the underlying complete formalisms.

\section{Universality vs. Effectiveness}
\setcounter{equation}{0}

The universality and effectiveness are always contradictory in
logic. In a fundamental theory such as general relativity or
unified field theory, such contradictory becomes even more
serious. The more universal the postulates, the more invalid
messages they contain, but the more concrete the constraints, the
greater the risk of failure the framework takes.

To explain the accelerating expansion of the present universe, one
prescription is to modify the Einstein-Hilbert Lagrangian by
\begin{eqnarray}
f(R)=-2\La+R+\vep_1 R^2 +\vep_2 R^3+\cdots. \label{1.18}
\end{eqnarray}
In \cite{Padman1,Padman2}, shifting the emphasis from Einstein's
field equations to a broader picture of spacetime thermodynamics
of horizons leads to a series of field equations constructed by
different order combinations of metric and Riemann curvature,
which includes Lanczos-Lovelock gravity. Of course, these theories
are more general than the Einstein's equation $G_{\mu\nu} +\La
g_{\mu\nu}= \kp T_{\mu\nu}$. Such theories can never be cancelled
by experiments, because we can never perform an experiment with 0
error. When we get an empirical result with relative error
$10^{-8}$, one may argue $|\vep_n|< 10^{-9}$. On the contrary, the
gauge invariance of the particle models such as $SU(2)$, $SU(3)$
is strong and narrow constraints in logic, and some possibly
correct field equations without such symmetries are arbitrarily
excluded. As a working hypothesis for concrete calculation, it may
be right or not. However, as a general principle, the risk of
being disproved by more accurate experiments and replaced by more
general principles is large.

Then what is the ideal feature of the postulates in the unified
field theory or the final theory? Recently, there are more and more
philosophical considerations on this
issue\cite{Ellis3}-\cite{gu1},\cite{philo1}-\cite{philo10}. Many
scientists believe the Principle of Equivalence has a compelling
force, and which leads to the metric theory of gravity. However this
point of view does not shared by all physicists\cite{Synge,Will}.
The principle of equivalence once indeed was a good guidance to
introduce curved space-time, but it may be only an approximate
principle in strict sense, because the concept `mass of a particle'
is just a classical approximation, and the trajectory of a fermion
with nonlinear potential is not an exact geodesic\cite{gu1}. The
real power of general relativity comes from that the curved
space-time is much more general and natural than the rigid Minkowski
space-time. The lack of universal simultaneity strongly suggests
that the realistic space of the Universe should be a curved
hypersurface. On the contrary, the principle of covariance of the
physical laws indeed has an inviolable feature. So the ideal
postulates of the final theory should have such philosophical depth,
self-evident meanings and an excellent balance between universality
and effectiveness. The `Singularity-free' postulate also has such
feature, so it is hopeful to become a universal principle of
physical laws. According to this principle, we may delete most terms
from (\ref{1.18}) and reduce it to the Einstein-Hilbert formalism.

\section*{Acknowledgments}
The author is grateful to his supervisor Prof. Ta-Tsien Li and Prof.
Han-Ji Shang for their encouragement. The explanation for the 4-d
coordinate system and simultaneity is completed according to the
discussion with one referee.

\end{document}